\documentclass[preprint]{aastex62}

\usepackage{graphicx}
\usepackage{url}
\submitjournal{ApJ}
\graphicspath{{./}{figures/}}

\shorttitle{Using Auto-correlations to Detect Periodicities}
\shortauthors{Allian, Jain \& Hindman}

\begin{document}

\title{A New Analysis Procedure for Detecting Periodicities within Complex Solar Coronal Arcades}
\correspondingauthor{Farhad Allian}
\email{fallian1@sheffield.ac.uk}

\author[0000-0002-4569-0370]{Farhad Allian}

\author[0000-0002-0080-5445]{Rekha Jain}
\affil{School of Mathematics and Statistics, University of Sheffield, S3 7RH, UK}

\author{B. W. Hindman}
\affil{JILA, NIST and University of Colorado, Boulder, CO~80309-0440, USA}

\begin{abstract}
We study intensity variations, as measured by the Atmospheric Imaging Assembly (AIA) on board the Solar Dynamics Observatory (SDO), in a solar coronal arcade using a newly developed analysis procedure that employs spatio-temporal auto-correlations. We test our new procedure by studying large-amplitude oscillations excited by nearby flaring activity within a complex arcade and detect a dominant periodicity of 12.5 minutes. We compute this period in two ways: from the traditional time-distance fitting method and using our new auto-correlation procedure. The two analyses yield consistent results. The auto-correlation procedure is then implemented on time series for which the traditional method would fail due to the complexity of overlapping loops and a poor contrast between the loops and the background. Using this new procedure, we discover the presence of small-amplitude oscillations within the same arcade with 8-minute and 10-minute periods prior and subsequent to the large-amplitude oscillations, respectively. Consequently, we identify these as ``decayless" oscillations that have only been previously observed in non-flaring loop systems.
\end{abstract}

\keywords{Sun: activity - Sun: atmosphere - Sun: corona - Sun: flares - Sun: oscillations}

\section{Introduction} \label{sec:intro}

Solar coronal arcades consist of brightly illuminated arches of hot plasma referred to as coronal loops. The arcades can act as waveguides for magnetohydrodynamic (MHD) waves, and these waves are of particular interest due to their diagnostic value in estimating the magnetic properties of arcades through seismology. Theoretical studies by \cite{Edwin1983} and \cite{Roberts1984} have, so far, formed the basis for most models of wave propagation in solar coronal loops. These studies describe the propagation of magnetohydrodynamic (MHD) waves along {\sl straight} magnetic tubes. In particular, many previous seismic analyses have attributed the observed oscillations to the motions of fast kink waves. Theoretical studies of the wave propagation in arcades with {\sl curved} field lines, or loops, have been rarer. \cite{Smith1997} studied fast MHD waves in dense curved potential field loops. Their study focused on the leakage of such waves across field lines. \cite{Selwa2007} considered a similar curved-arcade loop model and concluded that the main source of the wave attenuation was through such leakage. Recently, \cite{HindmanJain2015,HindmanJain2018} have demonstrated that fast MHD waves can be fully trapped by the magnetic field in an arcade under fairly common circumstances. Thus, fast waves can form resonances and wave leakage is not necessarily an essential process. 

Since the advent of the Transition Region and Coronal Explorer (TRACE), wave propagation in coronal loops has been observed in the extreme ultraviolet (EUV) as the loops oscillate in response to the passage of transient MHD waves from nearby flares \citep{Aschwanden1999, Nakariakov1999,Li2017}. Such loops exhibit transverse standing oscillations with periods ranging from a few minutes to several tens of minutes. \cite{Aschwanden2002} investigated 17 events with TRACE data and concluded that most of the oscillating loops do not fit the simple model of kink eigenmode oscillations, but instead suggest that the oscillations are flare-induced impulsively generated MHD waves, which decay rapidly either due to damping or wave leakage. Such observed large-amplitude attenuation has been generally attributed to resonant absorption, a mode conversion process whereby energy is transferred from the global transverse waves to local Alfv{\'{e}}nic waves \citep[e.g][]{Goosens2002, Ruderman2002, HindmanJain2018}. An alternate theory has also been proposed that explains the rapid signal attenuation as an interference effect that occurs whenever wave packets propagate along a multi-dimensional waveguide \cite{HindmanJain2014}. 

With high-cadence data from  the Atmospheric Imaging Assembly (AIA) onboard the Solar Dynamics Observatory (SDO) \cite[see][]{Lemen2012}, it is now clear that multiple loops within a single magnetic arcade often oscillate jointly \citep{Schrijver2002, Verwichte2009}. \cite{JainHindman2015} have reported that small phase shifts exist between such co-oscillating loops and suggest that such shifts could be caused by a moving driver or by the excitation of fast MHD waves that propagate across field lines from one loop to the other.

More recently, a distinct type of oscillation has been reported that is not clearly connected to any impulsive driver \citep{Wang2012, Anfinogentov2013, Nistico2013}. These are low-amplitude oscillations and do not appear to exhibit a temporal decay. As such, some have called these oscillations \lq\lq decayless\rq\rq. \cite{Anfinogentov2015} conducted a statistical analysis of 21 non-flaring active regions in the 171 {\AA} bandpass of SDO/AIA in order to estimate the regularity of this phenomenon. The average amplitude in the loop displacement is estimated to be $0.17$ Mm, with periods ranging from $1.5$ minutes to $10$ minutes. The nature of the driver of these oscillations remains unknown and various models have been suggested. Noting that these low-amplitude oscillations have poor phase-coherence over long durations, \cite{HindmanJain2014} considered a stochastically driven model of a 2D waveguide representing the entire coronal arcade. The decayless oscillations were excited by a distributed and stochastic source and appeared as a series of interference patterns formed by a multitude of MHD waves traveling through the waveguide. \cite{Nakariakov2016} have suggested that the decayless oscillations suffer the same decay mechanism as the flare-induced waves but supergranulation acts as a stochastic source that replenishes the lost energy.

In this paper, we present a new analysis method that uses auto-correlations of the traditional time-distance images to extract properties of the wave-field within the coronal arcade. This method has the salutary feature that it can be successfully applied to loops and arcades for which the traditional time-distance method would fail because of poor image contrast. We first validate this new method before illustrating these advantages. To do this, we measure the period of coronal loop oscillations using both our new procedure and the traditional time-distance method. We then compare the parameters measured with the two methods. Finally, we demonstrate the utility of the auto-correlation method in the detection of an additional periodicity in the form of low-amplitude oscillations that exist both long before and after the flares.

The paper is organized as follows. In Section \ref{sec:data}, we describe the observational data and the chronological events that triggered the coronal loop oscillations. In Section \ref{sec:time}, we present the image processing used to generate standard time-distance images of the oscillations. Subsequently, we present the results of a traditional fitting of the oscillations of the loops. In Section \ref{sec:auto}, we describe our new auto-correlation procedure, compare its results to the traditional method, and present an application of our new procedure to data for which the traditional fitting method would fail. Finally in Section \ref{sec:dis}, we discuss the implications of our findings.

\section{Observational data} \label{sec:data}

We study coronal loop oscillations on the Southeastern limb using EUV images obtained by AIA/SDO with unprecedented spatial ($1$ pixel $\approx$ $0.6$ arcsec) and temporal ($12$ s cadence) resolutions on 2014 January 27. The arcade of interest belonged to a multi-polar active region (AR) NOAA AR11967, which was behind the limb at the time of flare activity, and emerged a day later exhibiting a sunspot. The flaring activity is believed to have originated near the old active region AR11944 (S09, L=101) (see http://www.aurora-service.eu). While the flare is visible in all six EUV wavelengths, the arcade was predominantly visible in the $171$, $193$ and $211$ {\AA}\ channels and appeared as a bundle of illuminated arched threads, we refer to as loops. The dataset was chosen due to the off-limb nature of the arcade, where the loops have a higher visibility contrast against the darkness of the background. For the entirety of our study, we examined 12 hours ($3600$ time frames) of EUV imagery, starting from 2014 January 26 20:00 UT and ending on 2014 January 27 08:00 UT.
\begin{figure}[t]
	\begin{center}
		\includegraphics[width=\textwidth]{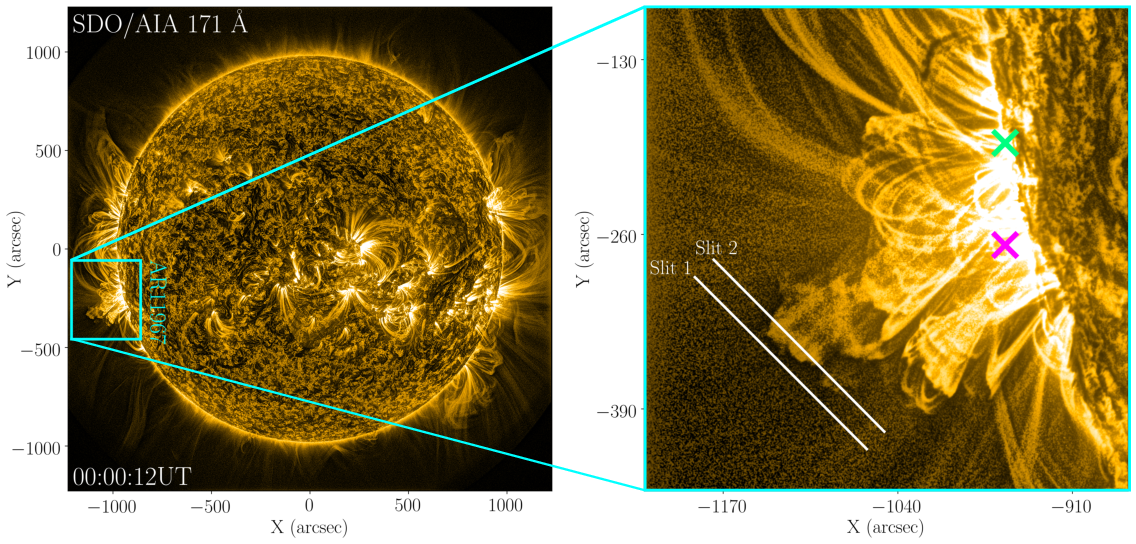}
		\caption{EUV image at the beginning of our dataset on 2014 January 27 00:00:12UT observed with SDO/AIA $171$ \AA. {\it Left panel}: Full-disk image indicating the active region of interest. {\it Right panel}: Zoomed-in view of the area contained in the box. The solid white lines correspond to a $~130$ Mm slit placed transverse to the apparent arcade. The magenta and green crosses correspond to the approximate positions of the flaring activity. The analyses of the loop oscillations are performed separately on each of these two numbered slits. Note that this image has been enhanced with the Multiscale Gaussian Normalization \citep{Morgan2014}. \label{Fig1} }
	\end{center}
\end{figure}
\setcounter{table}{0}

Figure \ref{Fig1} displays a EUV snapshot of the coronal arcade above NOAA AR11967 observed through AIA $171$ {\AA}. The loops in the arcade were seen oscillating around the time of two consecutive M1 class flares, which were behind the Southeastern limb and were recorded in X-rays by the GOES instruments. We investigate, in detail, the oscillations as they manifested along the two slits indicated in the right panel of Figure \ref{Fig1}.

\begin{figure*}[t]
\begin{center}
\includegraphics[width=0.8\textwidth,height=0.3\textheight]{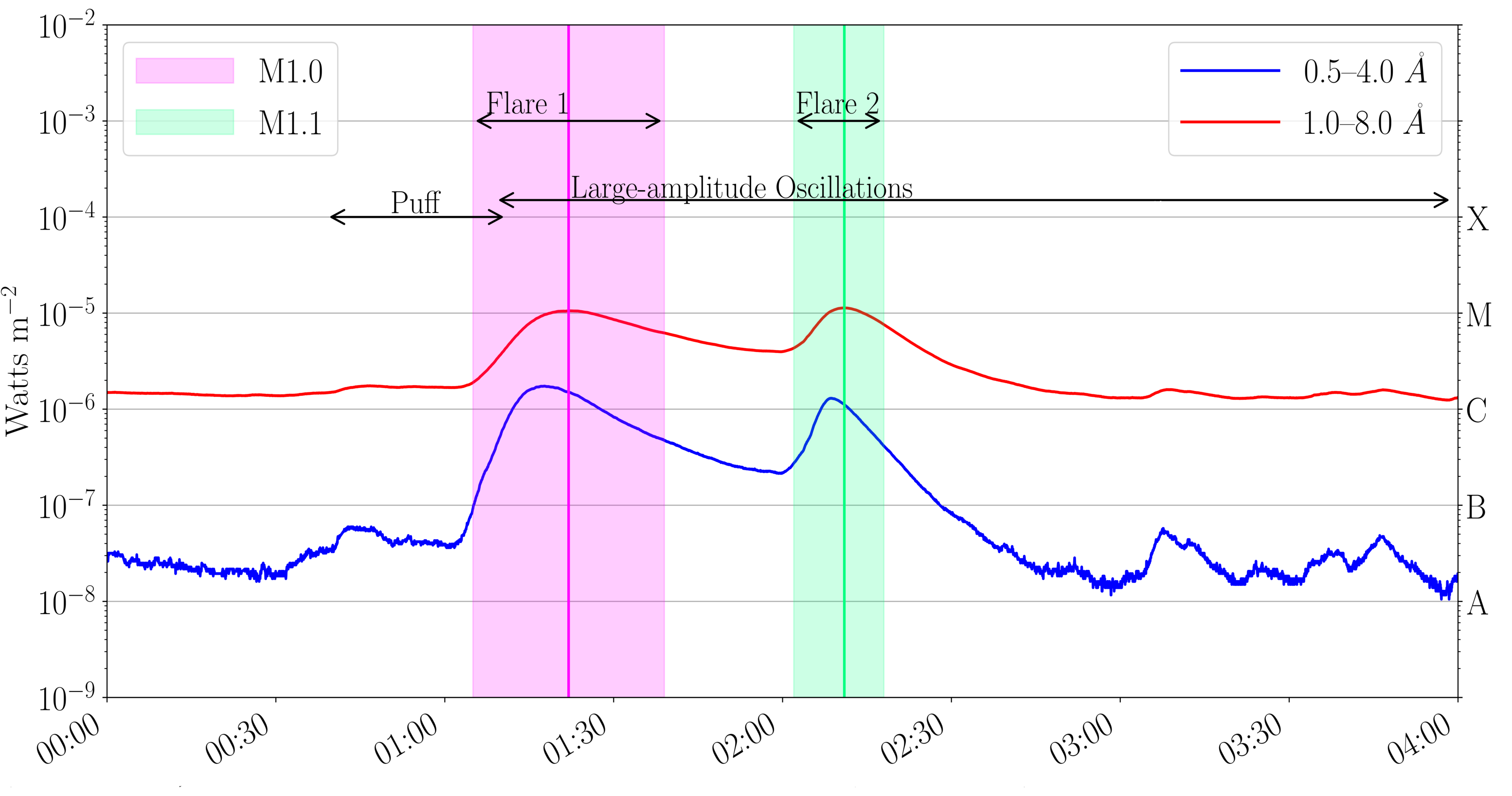}
\caption{Energy flux of the two M-class flares near NOAA AR11967 as detected by the GOES instrument. The magenta and green lines show the peak times of the first and second flares, respectively. The shaded areas correspond to the onset and final flare times. The duration of the puff and the coronal-loop oscillations are also indicated by arrows.\label{Fig2}}
\end{center}
\end{figure*}

Figure \ref{Fig2} shows the recorded X-ray flux by GOES in $4$\ {\AA} and $8$\ {\AA}. The first flare, located at latitude $16^\circ$ South and longitude $88^\circ$ East, was an M1.0 class flare, with a start time at 01:05 UT, a peak at 01:22 UT, and an end time at approximately 01:39 UT. After this an M1.1 class flare, at latitude $13^\circ$ South and longitude $88^\circ$ East, initiated at 02:02 UT, peaked at 02:11 UT, and ended at about 02:18 UT. Just before the first flare, a small wave-front or a puff-like structure (hereafter referred to as the ``puff") was also seen propagating away from the limb and throughout the arcade. Initially, the puff appeared near the limb around 00:40 UT, and became evident at 01:00 when it started moving. The initial motion of the puff from the flare site was visible in the AIA movies at all six EUV wavelengths. The life-time of the puff, as seen in 171 \AA\ passband movie, is also marked in Figure \ref{Fig2} with a double-headed arrow. A summary of the major events are shown in Table \ref{Table1} See \cite{Alzate2016} for details on coronal jets and puffs. Additionally, coinciding with the time of both flares, STEREO-B/SWAVES recorded two Type III radio bursts.
\begin{table}[h!]
	\renewcommand{\thetable}{\arabic{table}}
	\centering
	\caption{A chronological summary of the impulsive events that occurred in AR11967 as observed by AIA.} \label{Table1}
	\begin{tabular}{cccc}
		\tablewidth{0pt}
		\hline
		\hline
		Event & {Duration (UT)} & {Comments}\\
		\hline
		Puff &  00:55 - 01:05   & Visible in all EUV channels. \\
		1st Flare &  01:05 - 01:39     & M1.0 class.\\
		2nd Flare &  02:02 - 02:18    & M1.1 class. No puff observed.\\
		Large-amplitude oscillations &  01:10 - 04:00 & Predominant in $171$, $193$ \& $211$ {\AA}.  \\
		
		\hline
	\end{tabular}
\end{table}

\begin{figure*}
\begin{center}
\includegraphics[width=\textwidth,height=0.22\textheight]{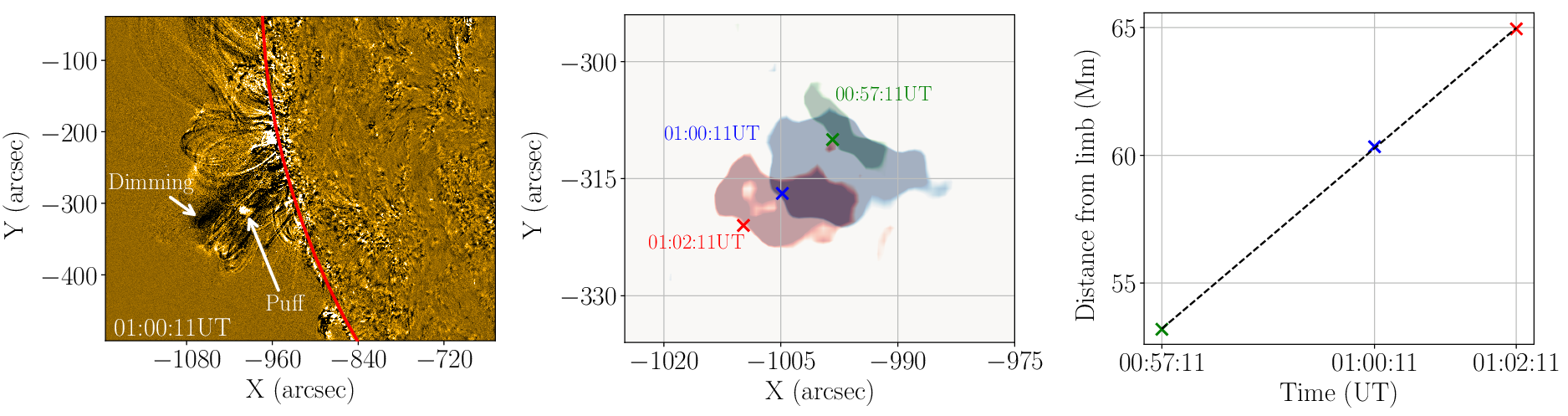}
\caption{{\it Left panel}: Base difference image revealing the puff before the onset of the first flare. The solar limb is indicated by the solid red line. {\it Middle panel}: Composite snapshots of the puff at three different times. {\it Right panel}: Projected initial distance of the puff from the solar limb as a function of time. \label{Fig3}}
\end{center}
\end{figure*}

After its initial stage of propagation, the puff was obscured by a bundle of several loops in the line-of-sight and so it was not possible to track it further. By carefully inspecting difference images, as shown in Figure \ref{Fig3}, we measure the distance the puff traveled from the limb at three different times (see the middle panel). We estimate the puff to have an initial projected propagation speed of about $40$ km s$^{-1}$ from the slope of the line shown in the right panel.

\section{Traditional Time-Distance Analysis} \label{sec:time}

%ANALYSIS

From the sequence of EUV images, it appears that the puff propagated away from the limb followed by the first M-class flare. The coronal loops above this AR were then seen oscillating during which a second flare occurred that did not excite further loop oscillations. In order to study the oscillations in greater detail, we extracted temporal intensity variations in the AIA $171$ and $193$ {\AA} channels along an approximately $130$ Mm long slit, indicated by Slit 1 in Figure \ref{Fig1}, and also shown by the white line in Figure \ref{Fig4} (left panel). The slit was placed perpendicular to the axis of the arcade and the time-distance images were created by temporally stacking the intensity along the slit at the AIA cadence. In order to remove small spatial-scale noise, we increase the signal-to-noise by applying a box-car smoothing of the intensity over a width of about $2$ Mm on either slide of the slit. The resultant intensity variations are shown in the right panels in Figure \ref{Fig4}, where the origin of the time-distance image corresponds to the bottom right point of Slit 1. The upper and lower panels corresponds to the 171 \AA\ and the 193 \AA\ bandpass, respectively.

%TIME-SERIES, METHOD 1
Further, to accurately extract the oscillatory properties with detail, the waveforms must be seen clearly with well-defined amplitude boundaries. In order to do this, the time-distance images were further enhanced by convolving the image with a weighted $3 \times 3$ kernel of the form:

\begin{equation}
    \left( \begin{array}{ccc}
    1 & 2 & 1 \\
    0 & 0 & 0 \\
    -1 &-2 & -1 \end{array} \right). \label{Eq1}
\end{equation}
This convolution performs a triangular temporal smoothing to the time-distance images and takes a numerical derivative with respect to the spatial position along the slit. As a result, the contrast is enhanced at the loop edges. Figure \ref{Fig5} presents the processed time-distance image for the 193 \AA\ bandpass, highlighting a variety of oscillating loops. These oscillations are also clearly visible in 171 \AA\ and 211 \AA\, but not in the 94 \AA\ and 131 \AA\ channels.

The dominant loop featured in Figure \ref{Fig5} undergoes a large-amplitude decaying oscillation. Interestingly, in addition to this, there were weaker small-amplitude oscillations that commenced near the onset of the second flare. These weaker oscillations appear near the bottom of the bundle of main loops. By fitting a Gaussian locally to the intensity of each pixel, we find the position of maximum brightness as a function of time and generate the time series. We fit the time series with appropriate sinusoidal functions of the form $\mathcal{A} \exp(-t/\tau) \cos(2\pi t/T + \phi) $, where $\mathcal{A}$ is an amplitude, $T$ is a period, $\tau$ is a damping time, and $\phi$ is a wave phase. The resultant fits are shown in Figure \ref{Fig5} as a sequences of red crosses and the fitting parameters are summarized in Table \ref{Table2}. To stabilize the fit of the small-amplitude oscillations, the decay rate $\tau^{-1}$ was fixed to zero.

\begin{table}[t]
\renewcommand{\thetable}{\arabic{table}}
\centering
\caption{Fitted parameters from Time-Distance Methods.} \label{Table2}
\begin{tabular}{ccccc}
\tablewidth{0pt}
\hline
\hline
Wavelength (\AA) & Amplitude (Mm) & Period (min.) & Damping time (min.)  & Phase ($^{\circ}$)\\ 
\hline
171 (large-amplitude)& (5.20 $\pm$ 0.75)  & (13.00 $\pm$ 0.06) & (34.43 $\pm$ 11.12) & (65.99  $\pm$ 3.73) \\
193 (large-amplitude) &(3.30 $\pm$ 0.76) & (13.02 $\pm$ 0.12) & (48.30 $\pm$ 2.70) & (-98.98 $\pm$ 6.57)  \\
193 (small-amplitude) & (0.68 $\pm$ 0.12) & (14.03 $\pm$ 0.21) & - & (20.38 $\pm$ 10.31) \\
\hline
\end{tabular}
\end{table}

\section{Auto-correlation Analysis} \label{sec:auto}

The intensity variations within traditional time-distance images contain an abundance of information. However, the standard time-distance analysis cannot capture this information when the loops are not well-defined, resulting in inaccuracies of the fitting parameters. In such a circumstance the loops cannot be fitted with fidelity and the method fails. In this section, we demonstrate the use of a new method that exploits auto-correlations of the raw time-distance images. We generate a 2D normalized auto-correlation function, which describes how well the image correlates with itself as it is shifted both temporally and spatially along the slit. As we will see, this procedure reveals the periodicities that remain hidden in the traditional time-distance analysis.

\begin{figure*}[t]
\begin{center}
\includegraphics[width=1\textwidth]{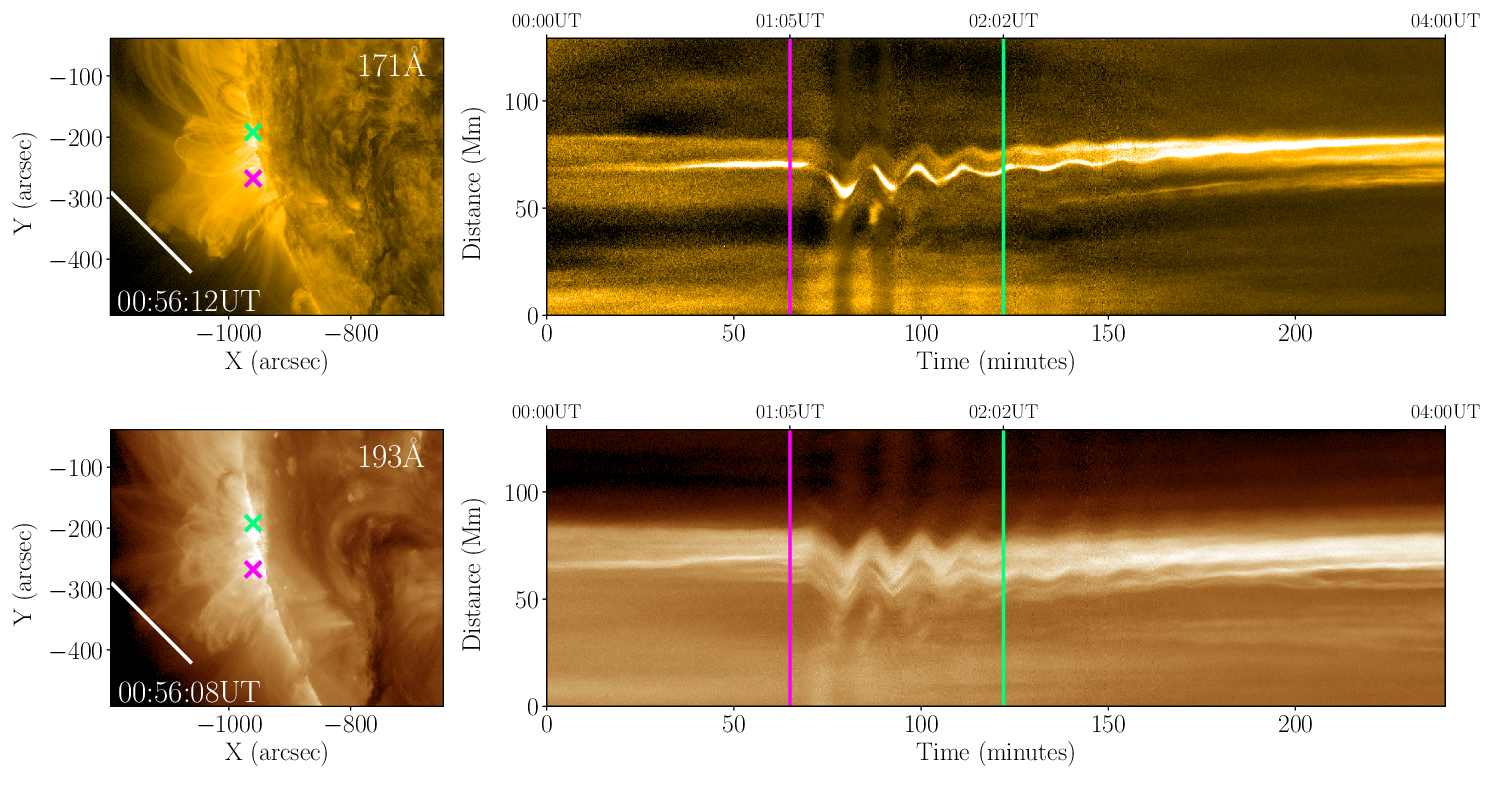}
\caption{Coronal arcades and oscillations. {\it Left panels}: EUV snapshots in the 171 \AA\ (top) and 193 \AA\ (bottom) channels, with Slit 1 indicated by the white line. The magenta and green cross indicate the onset times of the flares. {\it Right panel}: The corresponding intensity variations along Slit 1 in the aforementioned channels as a function of time. \label{Fig4} }
\end{center}
\end{figure*}

\begin{figure*}[t]
\begin{center}
\includegraphics[width=0.8\textwidth,height=0.37\textheight]{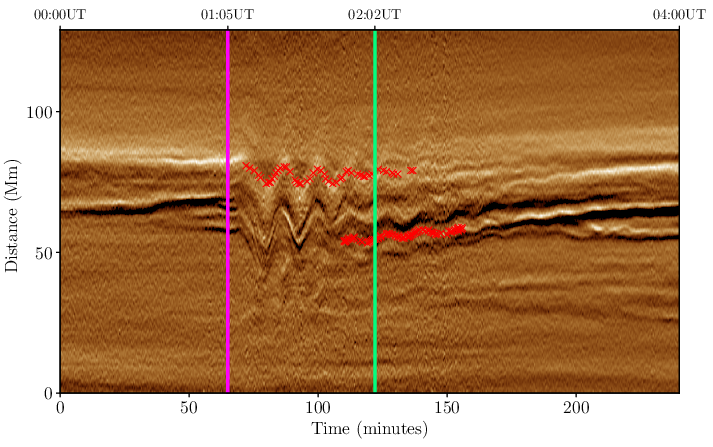}
\caption{Convolved time-distance image for the intensity variations as seen in the 193 \AA\ bandpass, shown in Figure \ref{Fig4}. The red crosses overplotted on top of the time-distance image are from the resultant time series fits.\label{Fig5} }
\end{center}
\end{figure*}

Many of the sharpest features in the time-distance images (as shown in Figure \ref{Fig4}) are fairly stationary as a function of time. Therefore, the application of a temporal high-pass filter will generate smooth images by removing a background. To accomplish this and to remove spurious signals from the steady features in each pixel, we compute a running-minimum difference image $\widetilde{I}(t,x)$ by subtracting a background intensity image $I_b(t, x)$ from the original time-distance images $I(t,x)$,

\begin{equation}
       \widetilde{I}(t,x) = I(t, x) - I_b(t, x) ,
\end{equation}

\noindent where $t$ is the temporal coordinate and $x$ is the spatial position along the slit. The background intensity $I_b$ is computed using the running-minimum difference suggested by \cite{Aschwanden2011} where $I_{b}(t_i,x) = {\min}[I(t_{i-k}, x),...,I(t_{i+k},x)]$. Specifically, $[t_{i-k},...,t_{i+k}]$ represents the time interval of length $2k$ placed symmetrically around the $i^{\text{th}}$ time frame. Here, after exploring a variety of values, we opt to use $k=5$ throughout the entire analysis.

The auto-correlation is defined in the standard manner as a function of spatial offset $\Delta x$ and time lag $\Delta t$,

\begin{equation}
	c(\Delta x,\Delta t) = \int \int \widetilde{I}(t,x) \widetilde{I}(x\pm\Delta x, t\pm\Delta t) \; dx dt .
\end{equation}

\noindent The integrals are approximated with discrete sums. We first zero-pad the time-domain signals and then generate a normalized auto-correlation $C(\Delta x,\Delta t) \equiv  c(\Delta x,\Delta t)/c(0,0)$. The periodic structures within the image are therefore revealed by a strong auto-correlation at the corresponding spatial offsets and time lags.


\begin{figure*}[t]
	\begin{center}
		\includegraphics[width=\textwidth]{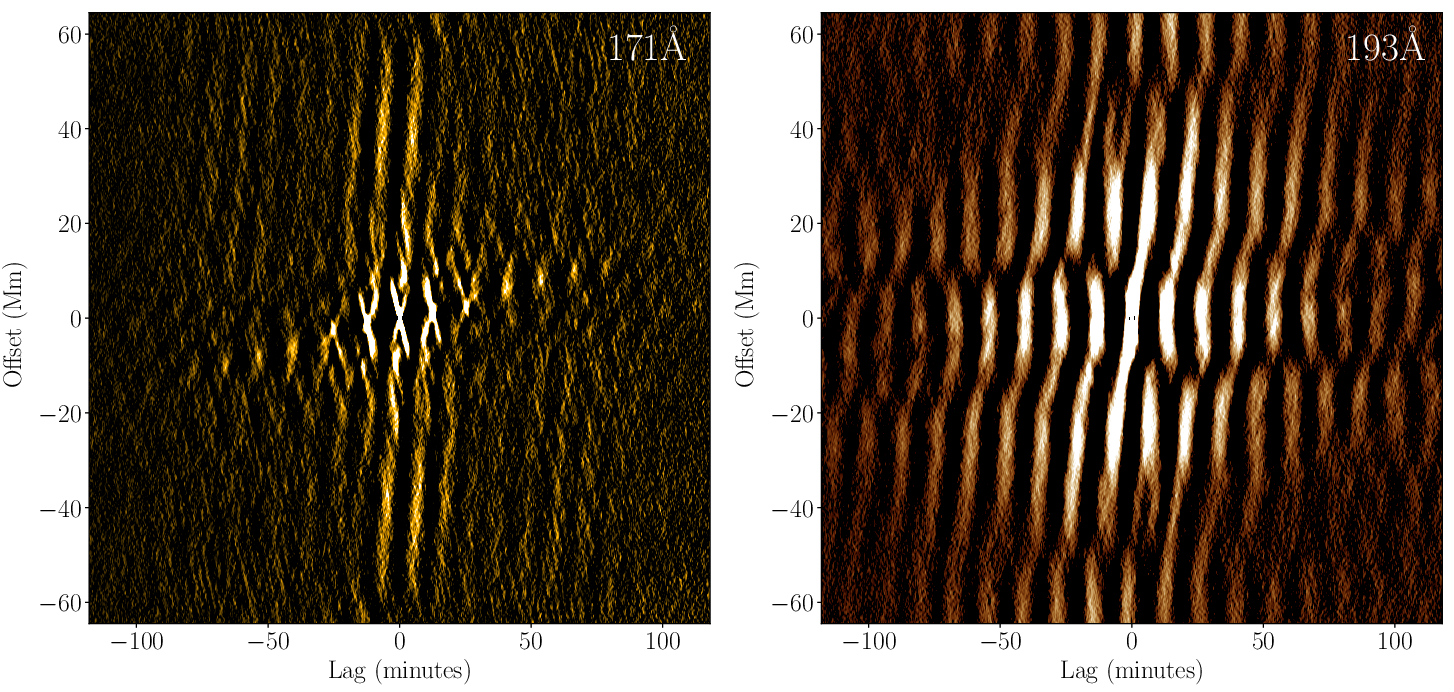}
		\caption{Auto-correlations of the time distance images generated as a function of spatial offset (Mm) and time lag (minutes) for Slit 1 in the 171 \AA\ (left) and 193 \AA\ (right) bandpass. \label{Fig6}}
	\end{center}
\end{figure*}

Figure \ref{Fig6} displays the auto-correlation as a function of spatial offset (measured in Mm) and time lag (measured in minutes) for the 171 \AA\ (left) and 193 \AA\ (right) wavelength bandpass. As expected, the maximum correlation occurs at zero time lag. The near vertical streaks have a slope due to the phase shift that exists amongst the multitude of oscillating loops sampled along the slit. Very noticeable X-like features are aligned in a sequence across time-lag with a different shallower slope. These Xs are most prominent in the 171 \AA\ channel and nearly invisible in the 193 \AA\ bandpass.

\begin{figure*}
	\begin{center}
		\includegraphics[width=\textwidth]{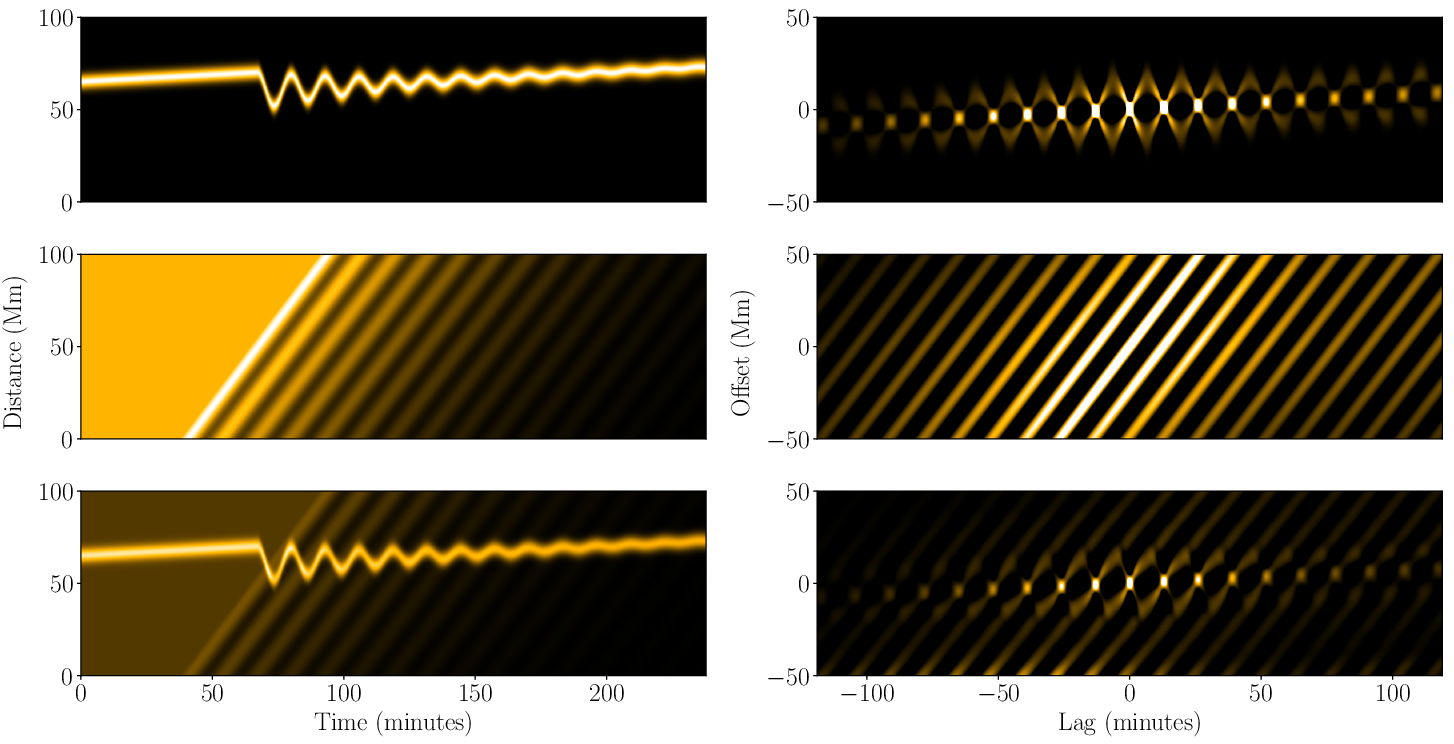}
		\caption{Artificial time-distance images (left) and their auto-correlations (right). {\it Top Panel}: Bright loop that suddenly begins oscillating, undergoes temporal decay, and slowly drifts along the slit. The auto-correlation of the bright loop with itself reveals a set of Xs whose centers are sloped according to the linear drift of the loop along the slit. {\it Middle Panel}: Background of faint loops that begin oscillating at different times thus introducing a phase shift that varies along the slit. The auto-correlation of the bundle of faint loops reveals tilted streaks whose slopes are fixed by the spatially varying phase between the different slit positions. {\it Bottom Panel}: Bright and faint-loop background superimposed. The prominence of the Xs depends on the relative brightness of the bright loop to the bundle of faint loops.\label{Fig7}}
	\end{center}
\end{figure*}

In order to understand the origin of the various slopes and features evinced by these auto-correlations, we created a synthetic dataset that consists of a bright oscillating loop embedded in a background of fainter dispersed loops that are also oscillating. The upper-left panel of Figure \ref{Fig7} shows the time-distance image of the bright loop in isolation. This bright loop starts oscillating, decays rapidly, and slowly drifts upward along the slit as time passes. The upper right panel reveals the 2D auto-correlation of this bright loop. We immediately see that the X-like structures that we observed in the auto-correlation of the coronal imagery is due to a bright oscillating loop correlating with itself. Further, the slope in the line passing through the centers of the sequence of Xs is caused by the temporal drift of the loop along the slit, possibly due to a moving driver or the puff.

The middle-left panel of Figure \ref{Fig7} shows the background of dispersed faint loops and the middle-right panel shows their auto-correlation. These figures make it clear that the tilted vertical streaks in Figure \ref{Fig6} are due to the bundle of faint loops inherent in the background of the image. The slope of the streaks arises from a phase shift between these loops in the bundle where the phase changes slowly along the slit. Finally, in the lower-left panel of Figure \ref{Fig7} we show the superposition of the time-distance image for the bright loop and the bundle of faint loops. The corresponding auto-correlation in the lower-right panel demonstrates that the prominence of the X-like features depends on the relative brightness contrast between the bright loop and the faint loop background and the relative phase shift. Note, in Figure \ref{Fig4} there is a clear bright loop in the 171 \AA\ channel (upper-right panel) with a well-defined amplitude and period, however, due to the relative contrast and the presence of a background of multiple faint loops, the same bright loop viewed in the 193 \AA\ bandpass appears only marginally brighter (lower-right panel). For this reason, the sequence of Xs is not so obvious in the auto-correlation of the time-distance image of the 193 \AA\ bandpass.

\subsection {Comparison of the Two Methods}

The auto-correlation contains a plethora of information about the oscillations, e.g., the phase coherence of the primary oscillations over multiple periods, whether the oscillation periods drift with time, phase relations between different oscillating structures, etc. However, at the moment, we will extract only the period of the dominant periodicity so that we can verify that our auto-correlation procedure and the traditional time series fitting method generate consistent results.

\begin{figure*}
\begin{center}
\includegraphics[width=0.9\textwidth]{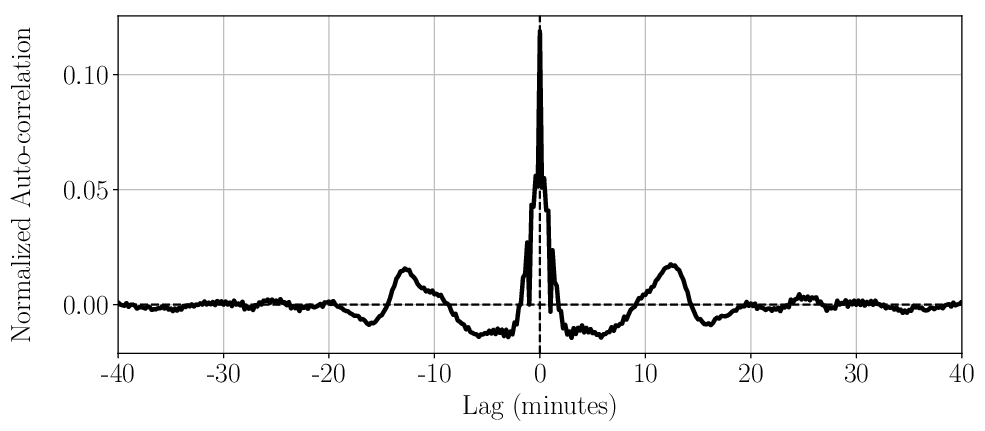}
\caption{Normalized auto-correlation function averaged over a narrow range of spatial offsets (between -3.5 Mm and 3.5 Mm) and plotted versus time lag. Each X-like feature in Figure \ref{Fig6} (left panel) produces a peak of positive correlation. The time lag of peak correlation for the first set of side lobes (to the right and left of the central correlation) corresponds to the dominant period of oscillation.\label{Fig8} }
\end{center}
\end{figure*}

The dominant period of oscillation can be obtained by measuring the location of the peaks (centers of the Xs) in the auto-correlation immediately to the left and right of the central peak at a time lag of zero (see Figure \ref{Fig6}). To illustrate these peaks, we average the auto-correlation over a band of spatial offsets, 
$\Delta x \in [-3.5 {\rm Mm}, 3.5 {\rm Mm}]$. This average is performed separately at each time lag $\Delta t$ and the result is shown in Figure \ref{Fig8}. The central peak arises from the correlation of the signal with itself at the same time. The peaks to the right and left come from correlating the current period with either the previous or the following period in the oscillation. Thus, the auto-correlation peaks at a time lag that corresponds to the wave period.

By measuring the time lag of maximum correlation for the first side peaks we deduce that the dominant oscillation within Slit 1 has a period of ($12.5$ $\pm\ 0.1$) minutes, a number consistent with the 13 minute period measured using the traditional method. If the oscillations were long-lived with steady periods, we would expect to see peaks at each multiple of the period as well and the amplitude of each peak would attenuate only slowly with time lag. That is clearly not the case here. We do see enhanced correlation at 25-26 minutes but the auto-correlation value drops rapidly from peak to peak, as one would expect for a rapidly decaying oscillation for which the higher multiples have fewer and fewer periods over which to correlate.

\subsection {Application of the Auto-correlation Method to Complex Bundles of Loops}

One of the advantages of the auto-correlation procedure is its ability to analyze loop systems for which the standard fitting method would fail. In particular, the auto-correlation procedure presented here can analyze bundles of loops which are poorly differentiated and criss-cross each other. As an illustration, we will analyze the intensity variations on Slit 2 which samples the arcade closer to limb. On this slit many loops co-exist in a complicated overlapping pattern. Further, we will analyze oscillations in periods that lie well before the initiating flares and well after. The goal is to seek low-amplitude ``decayless'' oscillations that would be too weak to otherwise fit.

\begin{figure*}
	\begin{center}
		\includegraphics[width=0.7\textwidth]{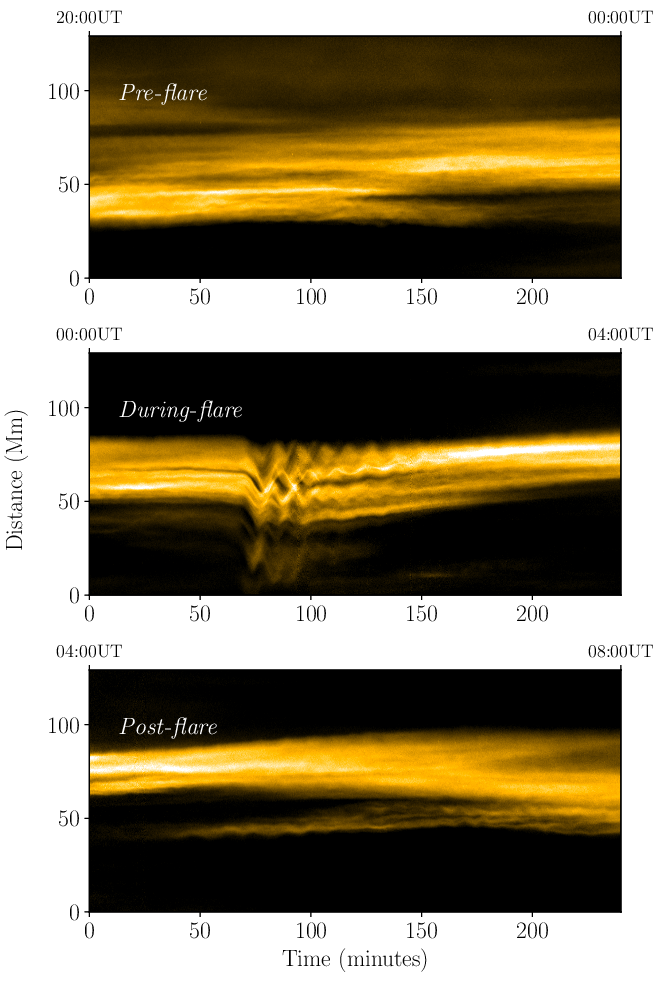}
		\caption{Intensity variations in 171 \AA\ bandpass as observed on Slit 2 closer to the limb. {\it Top Panel}: The time-distance diagram for a {\it pre}-flare phase (20:00--24:00 on 2014 Jan. 26). {\it Middle Panel}: Intensity variations during a phase coeval with the flares (0:00--04:00 on 2014 Jan. 27). This temporal window is identical to the one used to analyze the oscillations on Slit 1 (see Figure \ref{Fig4}). {\it Bottom Panel}: The time-distance diagram for a {\it post}-flare phase (04:00--08:00 on 2014 Jan. 27). Small-amplitude ``decayless'' decayless oscillations are present before and after the flares, but their displacements cannot be fitted due to the complex structure of the loop bundle. During the flaring phase, large-amplitude flare-induced oscillations exist, but once again the overlapping loops makes the fitting of the those loops problematic.\label{Fig9}}
	\end{center}
\end{figure*}

In Figure \ref{Fig9}, we show the intensity variations as they appeared on Slit 2 in AIA 171 \AA\ (see Figure \ref{Fig1} for the location of Slit 2). The upper panel shows the time-distance diagram for a {\it pre}-flare phase that spans the 4 hours immediately prior to the analyses presented previously in this paper (20:00--24:00 on 2014 Jan. 26 ). The middle panel displays the flaring phase (0:00--04:00 on 2014 Jan. 27), which identically matches the 4-hour period that was previously examined in detail for Slit 1. The bottom panel presents the {\it post}-flare phase which is the subsequent 4-hour period (04:00--08:00 on 2014 Jan. 27). Recall that the duration of flare activity reported by GOES was from 01:05 to 01:39 UT on 2014 Jan 27. Note, weak oscillation signatures (due to ``decayless oscillations'') do appear in the pre-and post-flare phases. However, the loop structure is so complicated and the oscillations so weak that it is not possible to extract the oscillation parameters by the standard fitting procedure as outlined earlier in Section \ref{sec:time}.

\begin{figure*}
	\begin{center}
		\includegraphics[width=0.65\textwidth]{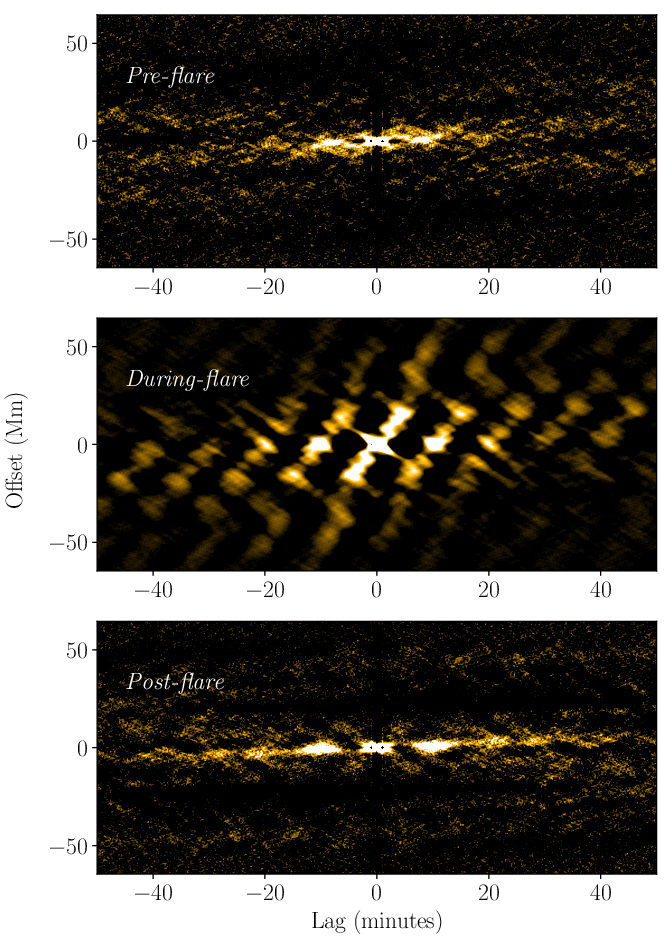}
		\caption{2D auto-correlations of time-distance diagrams appearing in Figure \ref{Fig9} obtained for Slit 2. {\it Top Panel}: The auto-correlation for the pre-flare phase. {\it Middle Panel}: The auto-correlations for the flaring phase (from 00:00 to 04:00 UT on 2014 Jan. 27). {\it Bottom Panel}: Auto-correlation of the post-flare phase (from 04:00 to 08:00 UT on 2014 Jan. 27). All are for the 171 \AA\ bandpass. The pre-flare and post-flare images reveal the existence of decayless oscillations whose lack of long spatial correlations indicate that different loops oscillate incoherently. Further, the presence of only one or two side-lobes on each side of the central correlation indicate poor phase coherence with time. \label{Fig10}}
	\end{center}
\end{figure*}

Figure \ref{Fig10} shows the auto-correlation functions derived from the three time durations indicated in Figure \ref{Fig9}. The top, middle, and bottom panels correspond to the {\it pre}-flare phase, flaring phase, and {\it post}-flare phase, respectively. The oscillations illustrated in the middle panel correspond to the same flare-induced oscillations discussed previously, but viewed at a position closer to the limb. There is a primary period and multiples of that period. All of the loops along the slit oscillate in concert, but do so with a phase shift that changes roughly linearly along the slit (as discussed in Section \ref{sec:auto}). Furthermore, the lack of clear Xs indicates that the entire bundle of loops contributes without a single dominant loop.

In the pre- and post-flare duration (top and bottom panel of Figure ~\ref{Fig10}), we see that the correlation has little signals for spatial offsets of much more than 5 Mm. There is a slight slope to the correlation that corresponds to a drift of the loop system along the slit as time passes. However, the concentration of signal near a spatial offset of zero indicates that loops do not correlate well with each other. We do, although, find that the temporal correlation possesses structure. For the pre-flare phase we see a central lobe at zero time lag and a single obvious side-lobe to each side of the central lobe. For the post-flare phase, there are additional side-lobes located at multiples of the time lag of the primary side-lobes. The simplest interpretation of these observations is that there is a primary frequency of oscillation at which each loop oscillates.  However, different loops along the slit lack coordination and oscillate at essentially random phases, suggesting that the temporal phase coherence of these small-amplitude oscillations are poor. In the pre-flare phase the amplitude is sufficiently low that only correlations with the immediately preceding or following phase is possible before noise (or another periodicity) dominates. In the post-flare phase the amplitude is larger, and we can see correlations arising from shifts of two or more wave periods.

\begin{figure*}
	\begin{center}
		\includegraphics[width=\textwidth]{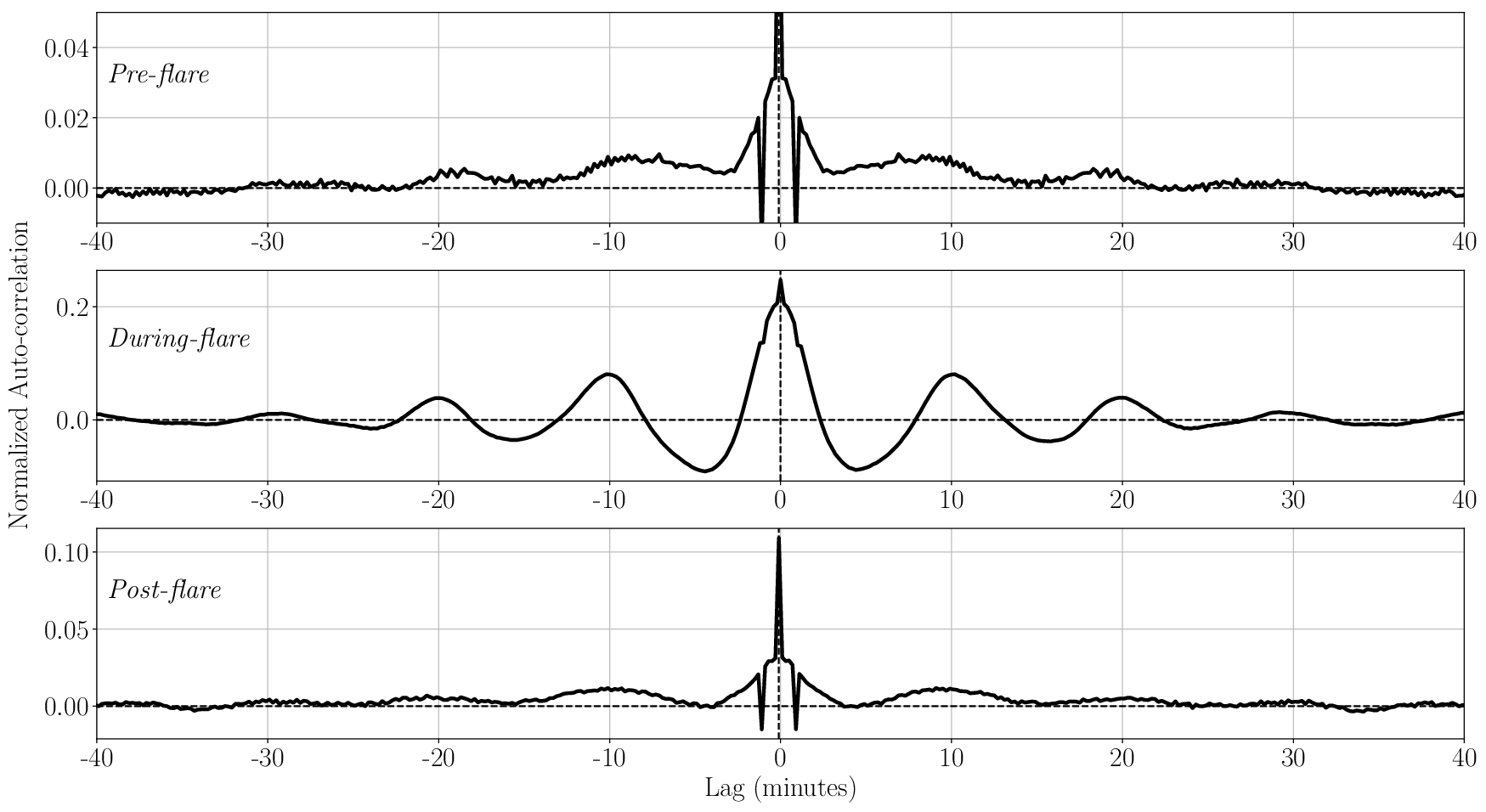}
\caption{Normalized auto-correlation function generated in the same manner as Figure \ref{Fig7}. The auto-correlations used to generate each panel are those shown in Figure \ref{Fig10}. {\it Top Panel}: Average correlation for the pre-flare phase. {\it Middle Panel}: Average correlation for the flaring phase and {\it Bottom Panel}: Correlation for the post-flare duration. \label{Fig11}}
	\end{center}
\end{figure*}

The exact periods of oscillation can be extracted from line plots of the auto-correlation at zero spatial offset. Averaging over a width containing the Xs, one obtains the correlations shown in Figure \ref{Fig11}. The dominant period of oscillation in the flaring phase is (10.4 $\pm$ 0.1) minutes and (9.8 $\pm$ 0.1) minutes in the post-flare phase. The dominant period is slightly shorter in the pre-flare phase at (8.1 $\pm$ 0.2) minutes.

%\section{Discussion} \label{sec:discussion}
\section{Discussion} \label{sec:dis}

\subsection{Oscillations During the Flares}

The primary period of oscillation of the flare-induced waves is clearly a function of the position of the slit. The oscillations observed on Slit 1, the slit furthest from the limb, possessed a dominant period of 12.5 minutes, while the slit closer to limb, Slit 2, had a shorter period, 10.3 minutes. Without performing similar analysis along a plethora of slits, we cannot ascertain whether the dominant period is a smooth function of height above the limb.

\cite{HindmanJain2015} have argued that the bundles of magnetic loops in an arcade oscillate together and that the true cavity is multi-dimension as opposed to an individual loop. Figures \ref{Fig4} and \ref{Fig5} clearly suggests this. Loops at different positions along the slit oscillate with phase shifts relative to each other and those phases change roughly linearly along the slit. If the waves propagate both along the field and transverse to the field, then we must entertain the possibility that magnetic pressure also plays a role, even if magnetic tension is the main restoring force.  The motions studied here, are clearly transverse to the magnetic field lines and the phase shift appears to be traveling across the magnetic field lines. This suggests that there is a compression of magnetic field lines, perhaps, indicating the presence of fast MHD waves.  

To test this idea further, we measure the speed of phase propagation along the slit directly from the slope of the near-vertical streaks in the auto-correlation diagrams. Doing so produces a phase speed of 140 km s$^{-1}$ for the 171 \AA\ bandpass and about 62 km s$^{-1}$ for the 193 \AA\ bandpass. Both slits generate similar phase speed values. These speeds are sufficiently slow compared to the local Alfv\'en speed that if the phase shift was caused by cross-field propagation the wave vector would need to be strongly radial, with only a small component aligned with the slits. 

An alternate interpretation for such a phase shifts is a moving driver. Recall that the initial propagation speed of the puff was 40 km s$^{-1}$ as projected on the plane of the sky. The similarity of this speed with the speed of phase propagation may be coincidental but it may also suggest that the puff acted as a moving wave-excitation source. This conclusion is further supported by the observation that the large-amplitude flare-induced oscillations appear to have been excited slightly before the occurrence of the peak of the x-ray flux from the first flare was recorded by the GOES observatories.

\subsection{Oscillations Before and After the Flares}

Along Slit 2, we see oscillations before, during, and after flaring activity. Prior to the flares the dominant period appears to be 8 minutes and each strand of the bundle of loops appears to oscillate rather incoherently with more distant strands. Well after the flares, the primary period appears to be 10 minutes and, once again, the individual loops that are well-separated oscillate somewhat incoherently.  The correlation length in both cases is roughly 5 Mm. We re-iterate and emphasize that these low-amplitude oscillations would be impossible to fit with the traditional time-distance method. The auto-correlation method presented is a promising tool for analyzing low-amplitude waves in coronal arcades at all times. A detailed study of the prevalence of decayless oscillations is now possible since we can easily study such waves even in complex arcades.

It is possible that the difference in dominant period between the pre-flare and post-flare phases (8 min. and 10 min. respectively) indicates a change in the arcade's resonant structure. The heating induced by the flares may have initiated a change in Alfv\'en speed along the loops under analysis. We note, however, that a similar change in apparent frequency may arise from a change in the distribution of waves with different cross-field wavenumber. \cite{HindmanJain2015,HindmanJain2018} have demonstrated that coronal arcades can act as waveguides, with resonances only in the radial direction (radial to the limb) and in the direction parallel to the field lines. The direction along the axis of the arcade (in this case parallel to the slit), may be unquantized. Each axial wavenumber has a different frequency and when energy is redistributed among this continuum of wavenumbers, the distribution of energy amongst modes with different periods is changed. One piece of evidence that supports this latter scenario is that the large-amplitude oscillations that were initiated during the flaring activity also have a dominant period of 10 minutes along Slit 2. It could very well be that the post-flare phase is dominated by waves initiated during the flare that have decayed in amplitude. Thus, the intensity variations in the post-flare phase may be a superposition of oscillations of periodicities of 8 and 10 minutes. 
\\
\\                           
\noindent {\bf Acknowledgments}: This work has been supported by STFC (UK). FA is grateful for the STFC studentship. BWH also acknowledges NASA (USA) through grants NNX14AG05G, NNX14AC05G, 80NSSC18K1125, and 80NSSC19K0267. We are grateful for the use of SDO/AIA and GOES data. 

%References                                                       
\bibliographystyle{yahapj}
\bibliography{AJH_ApJ_revised4.bbl}

\end{document}